\documentclass[twocolumn,pra,superscriptaddress,groupaddress,showpacs,preprintnumbers,amsmath,amssymb]{revtex4}
\ifx\pdfoutput\undefined
\usepackage{graphicx}
\usepackage[dvips]{color}
\usepackage{epsfig}
\else
\usepackage[pdftex]{graphicx}
\usepackage{epstopdf}
\fi
\usepackage{dcolumn}
\usepackage{bm}

\begin{document}
\title{Vortex trapping in suddenly connected Bose-Josephson junctions}
\author{Parag Ghosh}
\affiliation{Department of Physics, University of Illinois, Urbana-Champaign, Illinois 61801, USA}
\author{Fernando Sols}
\affiliation{Departamento de F\'{\i}sica de Materiales, Universidad Complutense de Madrid, 28040 Madrid, Spain}
\date{\today}

\begin{abstract}
We investigate the problem of vortex trapping in cyclically coupled
Bose-Josephson junctions. Starting with $N$ independent BECs we
couple the condensates through Josephson links and allow the system to
reach a stable circulation by adding a dissipative term in our
semiclassical equations of motion. The central question we address
is what is the probability to trap a vortex with winding number $m$.
Our numerical simulations reveal that the final distribution of
winding numbers is narrower than the initial distribution of total
phases, indicating an increased probability for no-vortex
configurations. Further, the nonlinearity of the problem manifests
itself in the somewhat counter-intuitive result that it is possible
to obtain a non-zero circulation starting with zero total phase
around the loop. The final width of the distribution of winding
numbers for $N$ sites scales as $\lambda N^{\alpha }$, where $\alpha
=0.47\pm 0.01$ and $\lambda <0.67$ (value predicted for the initial
distribution) indicating a shrinking of the final distribution. The
actual value of $\lambda$ is found to depend on the strength of
dissipation.
\end{abstract}

\maketitle

In the past few years, experiments on annular Josephson tunnel
junctions in superconductors \cite{carmi,monaco} and Bose-Einstein condensates \cite{anderson,cornell} have tried to address the
role of non-adiabaticity in the spontaneous production of
topological defects, a question that has bearing on early-universe
cosmology \cite{kibble,zurek,achucarro,donaire}. While a first type
of experiments \cite{monaco} have used a temperature quench through
a second-order phase transition from a normal to a superconducting phase, a second type \cite{anderson,cornell} uses
interference between initially independent condensates as a
mechanism to trap vortices. In the case of superconductors the
Kibble-Zurek scaling law \cite{zurek} relating the probability to
trap vortices to the quench rate has been tested. Experiments
connecting the independent BECs have similarly tried to test the
role of the merging rate in determining the probability for
observing vortices in the final BEC. Motivated by these experiments
we have studied numerically the related problem of a ring-shaped
Bose-Josephson junction array. We would like to stress that, while
there are similarities between our initial conditions and those of
the aforementioned experiments, there are also qualitative
differences that will be discussed later. Nevertheless, it is quite
conceivable that our findings here can be tested in future
experiments with ultra-cold atomic gases \cite{amico}.

\begin{figure}[tbp]
\includegraphics[width=2.5in]{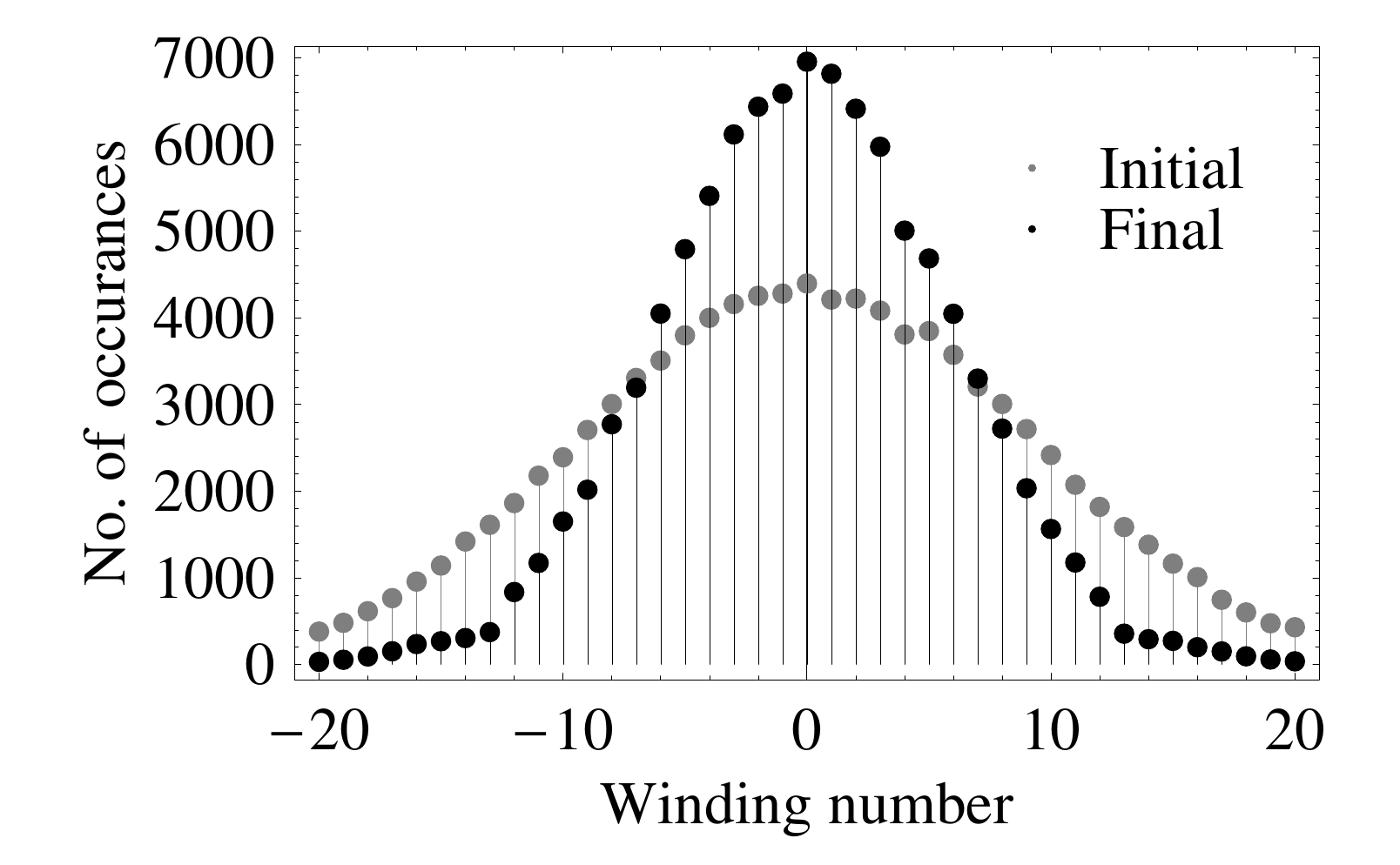}
\caption{Initial distribution of total phases and final distribution
of stable winding numbers for $N=10^3$ and $\protect\gamma =5$ for
$10^5$ runs } \label{histogram1}
\end{figure}

The problem we study here is that of $N$ independent Bose-Einstein
condensates which upon sudden connection become arranged on a ring
of weakly coupled condensates. We assume $r\leq \xi _{0}$, where $r$
is the single condensate radius and $\xi _{0}$ is the
zero-temperature healing length. This condition ensures that no
vortices form within the individual condensates, leaving us only
with vortices caused by the phase variation along the ring. At
$t=0$, simultaneous Josephson contacts are made between each
adjacent pair of condensates. As shown in Ref.\cite{zsl2} for the
case of two initially independent condensates, a relative phase is
quickly established once a few condensate atoms have hopped from one
side to another. Each pair of neighboring condensates behaves as if
a random relative phase $\varphi \in (-\pi ,\pi ]$ is chosen
locally. However, due to the single-valuedness of the macroscopic
wave function, there are only $N-1$ independent variables.
Therefore, in our simulations we choose  $N-1$ relative phases
independently, each following a flat distribution within the
interval $(-\pi ,\pi ]$ . The $N^{\text{th}}$ relative phase lies in
the same interval and is determined by the constraint that the total
phase variation around the ring should be $2\pi n$ ($n\in
\mathbb{Z}$). From the central limit theorem, we know that for
$N\rightarrow \infty $ the distribution of $n$  approaches a normal
distribution with FWHM = $2.354\,\sigma N^{1/2}$, where $\sigma
=1/\sqrt{12} $ is the standard deviation for a flat distribution in
the interval $(-\frac{ 1}{2},\frac{1}{2}]$. A key point is to
realize that the classically stable fixed points correspond to all
the relative phases being equal (modulo $2\pi $) to a value $2\pi
m/N$, where $m \in \mathbb{Z}$ is the winding number or charge of
the final vortex conguration. To allow our system to converge to one
of these fixed points we let each link follow a semiclassical
Josephson equation which includes a phenomenological dissipation
term characterized by a single parameter $\gamma $. Such dynamics
allows the system to go through phase slips at individual junctions.
Thus, generally $m\neq n$. A number of interesting results are
obtained:

%\begin{itemize}

%\item[1.]
(i) The distribution of the final winding number deviates from the
initial distribution for all values of $N$ and $\gamma $. That final
distribution for $m$ is narrower than the initial distribution for
$n$, indicating an increased probability for low-charge vortex
configurations (see Fig. \ref{histogram1}).

%\item[2.]
%II. For $\gamma \leq 3.0$ the width of the final distribution shrinks upon decreasing $\gamma$ (inset Fig. \ref{widthgamma}).

%\begin{figure}
%\includegraphics[width=2.5in]{final_histograms_gr13.pdf}
%\caption{\label{dvsgamma}Fitting plots like that of Fig. \ref{scaling} with $d \times N^{\alpha}$ one obtains $\alpha=0.47\pm 0.01$. The above figure shows how $d$ scales with $\gamma$. Note that for $\gamma<3.0$ the width decreases for decreasing $\gamma$. However for large $\gamma, d$  oscillates around a value less than 1, indicating the final distribution is always narrower than theinitial distribution irrespective of $\gamma$. Inset shows how the full width at half maximum (FWHM) of the final distribution of winding numbers scale with dissipation parameter $\gamma$ for $N=1000$.}
%\end{figure}

\begin{figure}[tbp]
\includegraphics[width=2.5in]{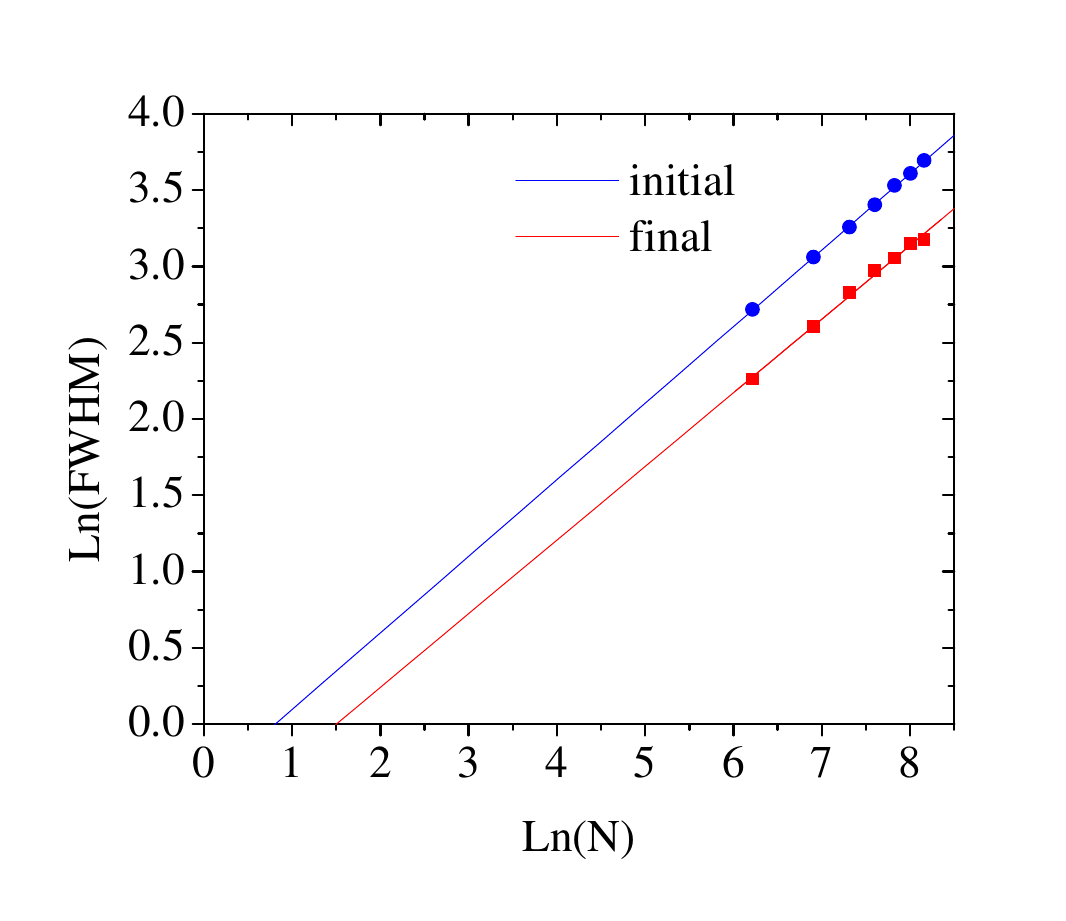}
\caption{Red plot shows how the FWHM of the final distribution of
winding numbers scales with $N$ for $\protect\gamma=6$. The scaling
exponent is $\protect\alpha=0.47\pm0.01$ and the prefactor
$\protect\lambda=0.55\pm0.05$. Blue plot shows the scaling of FWHM
of the initial distribution of total relative phases:
$\protect\alpha=0.50\pm 0.01$, $\lambda=0.67\pm 0.05$ (color
online). } \label{scaling}
\end{figure}

\begin{figure}[tbp]
\includegraphics[width=3in]{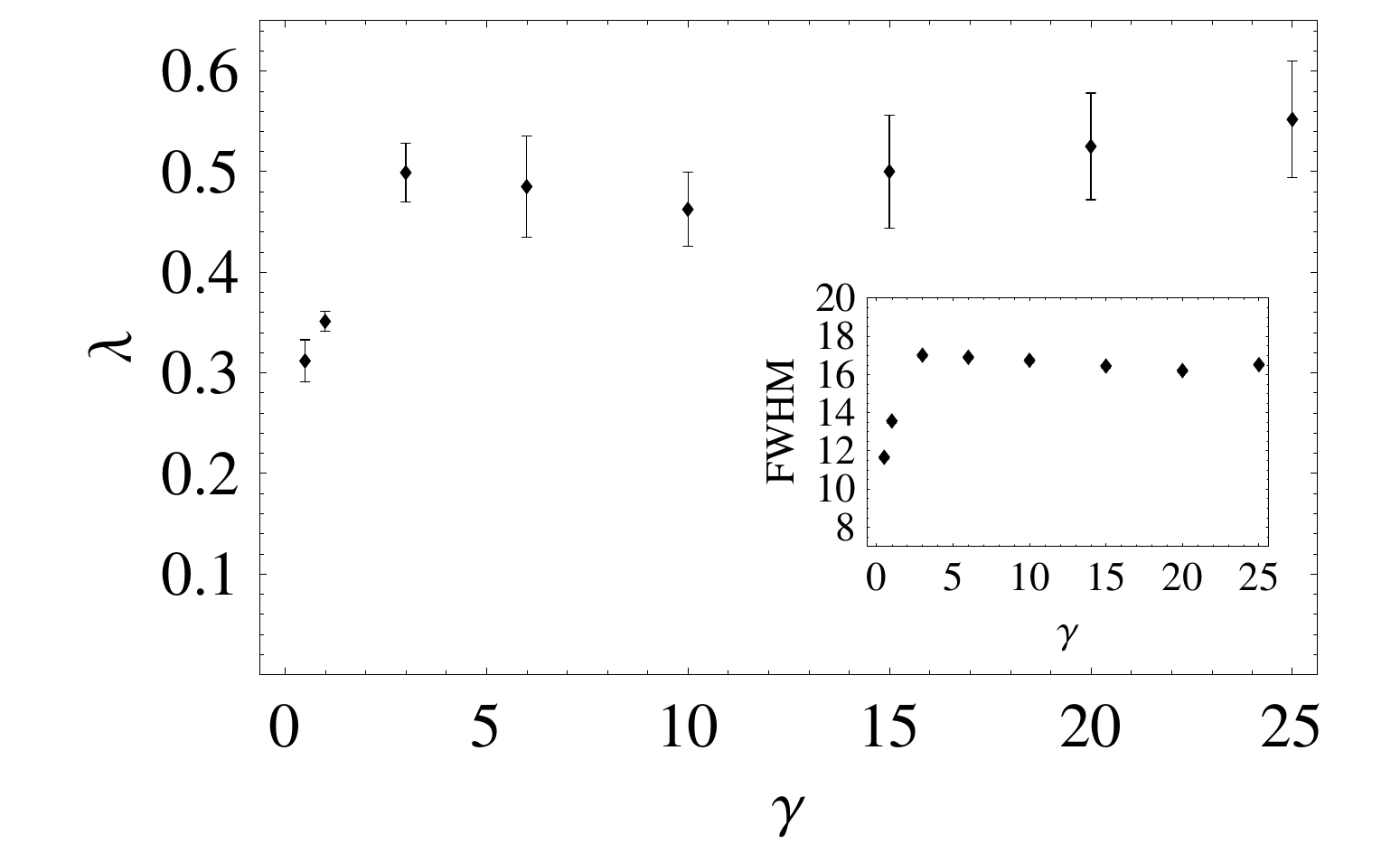}
\caption{Prefactor $\protect\lambda$ as a function of $\protect
\gamma$. Note $\protect\lambda<0.67$ for all values of
$\protect\gamma$. Inset shows how FWHM of the final distribution of
winding numbers scale with $ \protect\gamma$ for $N=10^3$.}
\label{dvsgamma}
\end{figure}

%\item[3.]
(ii) The width of the final distribution scales with the size of the
system as $\lambda N^{\alpha }$, where $\alpha =0.47\pm 0.01$,
independent of $\gamma $ and $\lambda <0.67$ (normal distribution
value), indicating a shrinking of the basins of attraction for
higher winding numbers (see Figs. \ref{scaling}, \ref{dvsgamma}).
For $\gamma \leq 3$ the width of the final distribution shrinks upon
decreasing $\gamma $ (see inset of Fig. \ref{dvsgamma}).

%\begin{figure}
%\includegraphics[width=2.5in]{final_histograms_gr12.pdf}
%\caption{\label{dvsgamma}Fitting plots like that of Fig. \ref{scaling} with $d \times N^{\alpha}$ one obtains $\alpha=0.47\pm 0.01$. The above figure shows how $d$ scales with $\gamma$. Note that for $\gamma<3.0$ the width decreases for decreasing $\gamma$. However for large $\gamma, d$  oscillates around a value less than 1, indicating the final distribution is always narrower than the initial distribution irrespective of $\gamma$.}
%\end{figure}

\begin{figure}[tbp]
\includegraphics[width=2.5in]{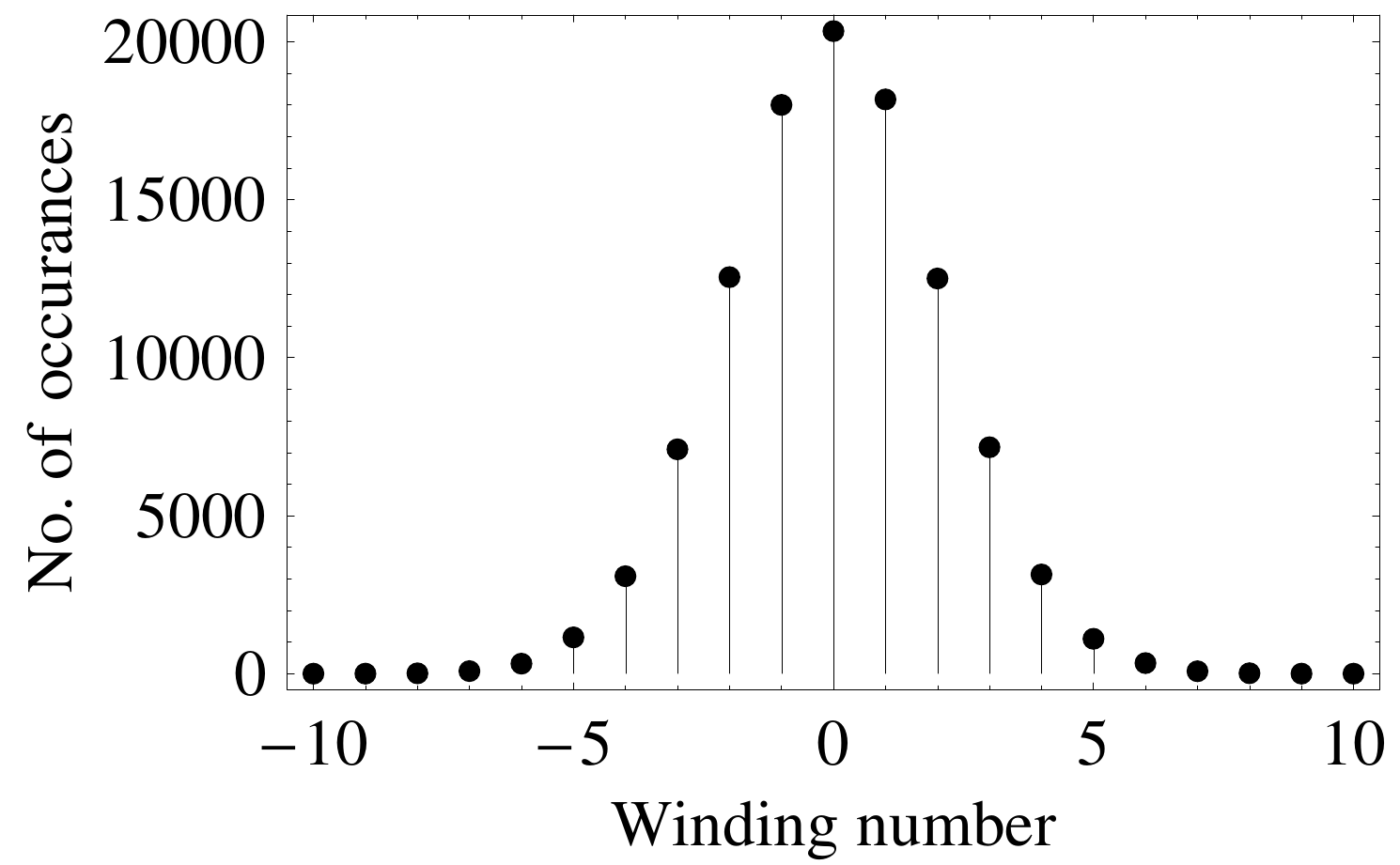}
\caption{Restricted to configurations $\sum_i \protect\phi
_{i,i+1}=0$, this histogram for final winding numbers shows that
even in the high friction limit one can obtain a non-zero
circulation. The above simulation uses $N=10^3$ and
$\protect\gamma=50$.} \label{conditional}
\end{figure}

%\item[4.]
(iii) If one focuses on initial configurations with $n=0$, the final
distribution of winding numbers in the limit of large $N$ is still a
Gaussian centered around $m=0$ with a nonzero spread (see Fig.
\ref{conditional}). This reflects the fact that a finite fraction of
the initial configurations with zero total phase have Josephson
coupling energies higher than those which correspond to nonzero
final winding numbers.

%\end{itemize}

We start our analysis of the Josephson dynamics by stating a
theorem: If $N$ BECs with random relative phases are coupled by a
nearest-neighbor Josephson coupling on a one-dimensional lattice
with periodic boundary conditions, a necessary condition to obtain a
metastable non-zero circulation of winding number $2\pi m$ is
$N>4m$, the case of $4m$ links being marginal. The proof is as
follows:

Let us assume that each Josephson junction is described by a
two-mode Josephson Hamiltonian:

\begin{equation}
H=-E_{J}\sum_{i}\cos {\phi _{i,i+1}}+(E_{C}/2) \sum_{i}n_{i}^{2}
\label{hamiltonian}
\end{equation}
where $E_{J}$ is the Josephson coupling energy, $E_{C}$ is the
charging energy, $\phi _{i,i+1}$ is the relative phase between $i$
and $i+1$ (with $i=N+1$ identified to $i=1$) and
$n_{i}=N_i-N^{(0)}_{i}$ is the deviation from the equilibrium value
at condensate $i$. We assume all $N^{(0)}_i$s to be the same and
initially $n_i=0$, so that $\sum_i n_i=0$ throughout the entire
evolution. In the classical limit this Hamiltonian can be mapped
into that of coupled rigid pendula, with the first term denoting the
\textquotedblleft potential energy" and the second term the
\textquotedblleft kinetic energy" of the pendula system. Now
consider a system with $N$ links and a total phase difference of
$2\pi m$ around the loop. As stated earlier, the fixed point
corresponding to a circulation of charge $m$ is given by the
configuration where all the phases are $\varphi _{m}=2\pi m /N$
(modulo $2\pi $). Hereafter, we simplify the notation $\phi_i \equiv
\phi_{i,i+1}$. To determine whether this fixed point is stable we
consider a configuration where $\phi _{i}=\varphi _{m}+\epsilon_i$
with $\sum_i \epsilon_i=0$ and $\epsilon_i \rightarrow 0$. The
potential energy of this new configuration with respect to the fixed
point is, up to second order in $\epsilon_i$, given by $\Delta
E(\epsilon_i)=(\cos {\varphi _{m}})\sum_i \epsilon_i ^{2}$. For the
fixed point to be stable we should have $\Delta E(\epsilon_i)>0$,
which requires $N>4m$. This theorem can equally be applied to a
system of $XY$ spins coupled by Heisenberg interaction. A corollary
is that final configurations satisfying $N/4\leq m\leq N/2$ are
unstable.

For a more generic analysis of the fixed points and their basins of
attraction we derive from Hamiltonian (\ref{hamiltonian}) a set of
semiclassical equations of motion for the relative phases and
currents at each junction:
\begin{equation}
\ddot{\phi}_{i}(t)=E_{C}\left[ 2j_{i}(t)-j_{i+1}(t)-j_{i-1}(t))
\right]   \label{phi}
\end{equation}
\begin{equation}
j_{i}(t)=-\sin {\phi _{i}(t)}-\gamma \dot{\phi}_{i}(t)  \label{n}
\end{equation}
Here time and energies are expressed in units of $E_{J}^{-1}$ and
$E_{J}$ ($\hbar=1$), respectively. It is important to note that for
cyclically coupled Josephson junctions the variable canonically
conjugate to, say, $\phi _{i}$ is not $ (n_{i}-n_{i+1})$ but rather
the quantity $\int_{0}^{t}j_{i}(t)dt$. We have also added a
phenomenological dissipative term of the form $-\gamma
\dot{\phi}_{i}$ in the equation of motion for $j_{i}$ while
neglecting finite-temperature noise. It is important to add this
term for the system to converge to one of the fixed points. From our
knowledge of three or more coupled pendula we know that the system
of equations (\ref{phi})-(\ref{n}) is chaotic \cite{nerenberg} and
without any damping would typically explore the whole phase space
without converging to a fixed point. To verify this point, we have
investigated the dynamics of Lyapunov exponents for the case of
$N=3$.  To ensure that the system is in the Josephson regime we take
$E_{C}/E_{J}=0.01$ in all our simulations. We find that 3 out of 6
Lyapunov exponents are positive, indicating chaotic behavior. We
note that the Ohmic nature of the dissipative term is only justifed
at high temperatures \cite{zapa98} or at low temperatures if each
condensate lives in a large box \cite{meie01}.

An interesting property of Eq. (\ref{phi}) is that $\sum_{i}\phi
_{i}$ is a mathematical constant of motion. However, physically the
system can still change its winding number by going through phase
slips at any junction. It will be useful to incorporate the above
constant of motion by imposing the restriction ${\phi _{i}\in (-\pi
,\pi]}$  only at $t=0$ and removing it for later times. Of course
the physical quantity which is observed at the end of the evolution
is the Josephson current at each junction, which depends on the
relative phase modulo $2\pi $.

In order to generate statistics, we consider a large number of
different initial configurations, with the relative phases and
numbers chosen as explained earlier.
%Also, we recall that the
%charging energy at each junction is chosen to be zero, $n_{i}(0)=0$.
Equations (\ref{phi})-(\ref{n}) are then numerically integrated for
each set of initial conditions. After the average current has
reached its final equilibrium value, its magnitude equals $\sin
(2\pi m/N)$ and the value of the final winding number $m<N/4$ is
uniquely extracted. A histogram is then plotted for all values of
$m$ and its width is recorded. To obtain the scaling law we have
calculated the width as a function of $N$ and fitted it to a
function of the form $\lambda N^{\alpha }$. The process is repeated
for different values of $\gamma$.

\begin{figure}[tbp]
\includegraphics[width=2.5in]{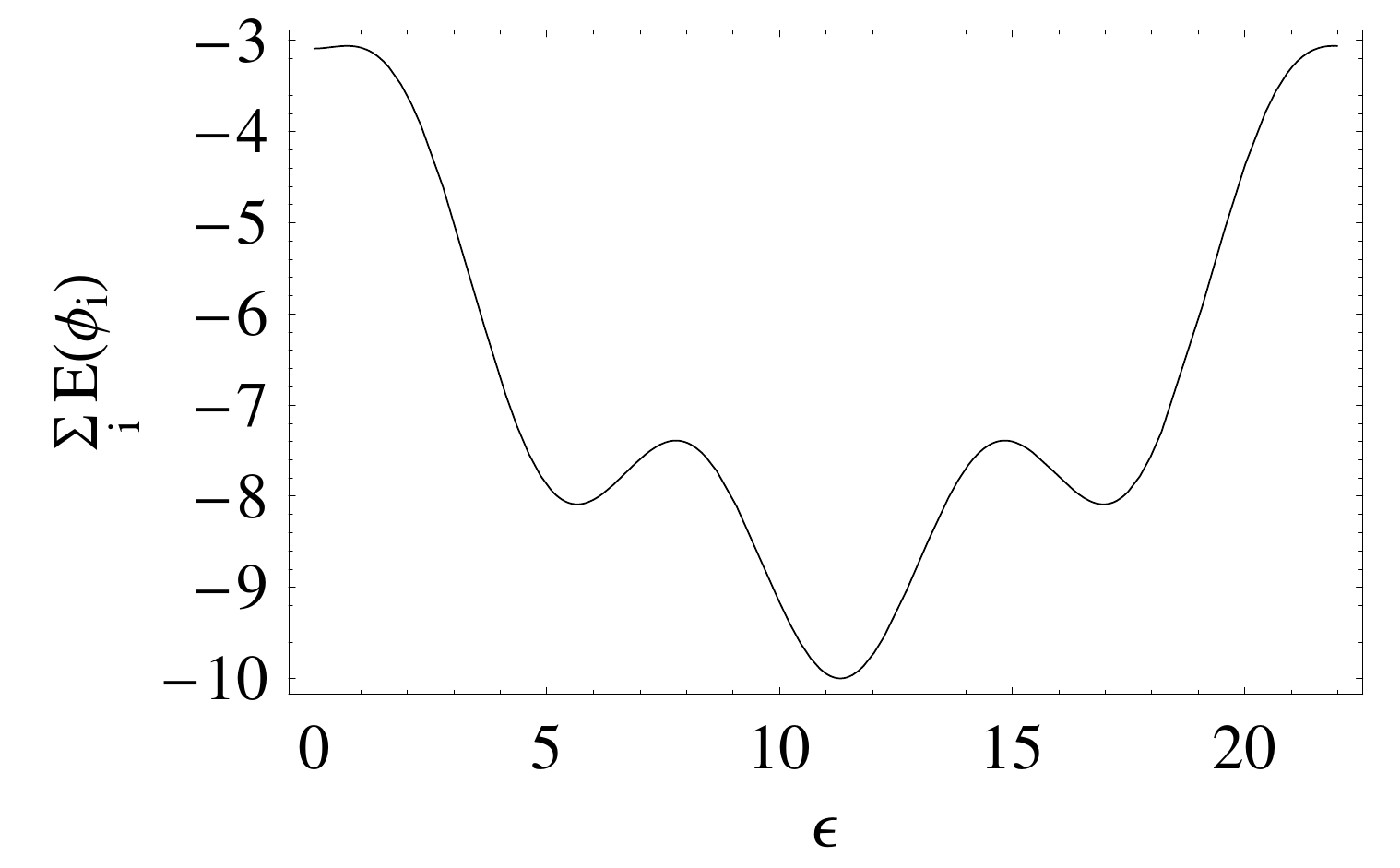}
\caption{Potential Energy landscape for $N=10$ and a certain class
of configurations:
$m=2;\,\protect\phi_{i}=4\protect\pi/10+\protect\epsilon
,\,\protect\phi_{j}=4\protect\pi/10-\protect\epsilon/9\,; \,j\neq
i$. Winding number zero is the global minimum of energy landscape
and here occurs at $\epsilon =3.6\pi$.} \label{landscape}
\end{figure}

To get a qualitative idea of the dynamics and the role of
dissipation, we consider a certain class of initial configurations
where $\phi _{1}=\varphi _{m}+\epsilon $ while
$\phi_{i}=\varphi_{1}-\epsilon /(N-1)$  for $2\leq i \leq N$. Given
$\phi _{1}$, this configuration has the lowest potential energy.
Fig. \ref{landscape} shows the potential energy for such a
configuration as a function of $\epsilon $ for $N=10$ and $m=2$. The
first minimum corresponds to the fixed point $K_{2}$ ($\phi
_{i}=\varphi _{2}$ for all $i$) followed by the fixed point $K_{1}$
($\phi _{1}=\varphi_{1}+2\pi\,;\,\phi_{i}=\varphi _{1}$ for all
$i>1$) and so on so forth. The global minimum of the energy
landscape is the configuration $K_{0}$ with zero winding number.
Starting with the initial configuration mentioned above, Fig.
\ref{landscape} shows the path of steepest descent from $K_2$ to
$K_0$. Starting from a local minimum one can characterize the size
of the basins of attraction by the value $\epsilon _{c}$ which
$\epsilon $ takes at the next nearest local maximum. However, one
should be warned that such an estimate applies only to the specific
class of initial configurations described above.

The role played by dissipation can also be elucidated by studying
that class of configurations. Suppose $\epsilon
>\epsilon_{c1}$, where $\epsilon_{c1}$ is the first critical value
of $\epsilon$. The system starts at an unstable point and as it
rolls down to the fixed point with one less winding number, loses
kinetic energy due to friction. If it arrives at the next stable
point with kinetic energy less than what is needed to overcome the
next barrier, then it settles down at the fixed point $K_{m-1}$.
However, if it has enough kinetic energy to roll over the next
barrier then the final winding number would be less than $(m-1)$. A
similar role can be envisaged for dissipation in the general
multidimensional landscape: For large $\gamma$, the system settles
down in the nearest valley; for small $\gamma$, the particle may
escape the initial basin and lower its winding number. Thus low
friction enhances, by a moderate factor, the probability of ending
in a low-charge configuration, as suggested by Fig. \ref{landscape}
and confirmed by Fig. \ref{dvsgamma}.

%Inspection of Fig. 5 also suggests a qualitative understanding of the effect of dissipation in the
%general case: The particle always starts with zero kinetic energy. In the limit of large friction, it lands in a typically close lake (fixed point) to which it is
%connected through the steepest descent path. For weak friction, and depending on the initial direction of motion, the particle may alternatively escape the
%initial basin and glide over the multidimensional landscape gradually lowering its energy to eventually land in a lower lake, i.e. in a lower-charge vortex
%configuration. Thus low friction enhances, by a moderate factor, the probability of ending in a low-charge configuration, as revealed by Fig. \ref{landscape}.
%Lastly, the reader
%is warned not to preclude, by noting the energy landscape (Fig. \ref
%{landscape}), the possibility to start with a given $m$ ($m=0$) and end up
%with a circulation of greater than $|2\pi m|$ ($m\neq 0$), a result (Fig.
%\ref{conditional}) that comes from more generic considerations.
%\begin{figure}
%\includegraphics[width=2.5in]{final_histograms_gr4.pdf}
%\includegraphics[width=2.5in]{final_histograms_gr5.pdf}
%\caption{\label{dissipation} The above two plot shows how the average current $\sum_i J_i(t)/10$ varies with time for $N=10$ and the same set of initial conditions with $\sum_i \phi_i=4\pi$ for two different values of $\gamma$. Note that when $\gamma=5.0$ the final current saturates to a value corresponding to winding number $2\pi$ and for $\gamma=0.5$ the same set of initial conditions yeilds no circulation.}
%\end{figure}

For a semi-analytical discussion of the basins of attraction we
focus on the case of $N=5$ (stable $m=0,\pm 1$) and high friction.
Let $P(m)$ be the probability of landing in a final vortex
configuration of charge $m$, $Q(n)$ the initial probability for
$\sum_i \phi_i=2\pi n$, and $P(m|n)$ the probability to obtain a
final charge $m$ conditioned to $\sum_i \phi_i=2\pi n$. Below we
estimate $P(1)$ and show that $P(1)<Q(1)$. First we note:
$P(1)=P(1|1)Q(1)+P(1|0)Q(0)+P(1|-1)Q(-1)$. We therefore begin by
estimating $P(1|1)$. The limit of high friction ensures that the
system follows the path of steepest descent towards the nearest
stable fixed point. The system always resides on the hypersurface
$S_n$ defined by the constant of motion $\sum_{i}\phi _{i}=2\pi n$.
Note that, on the surface $S_{1}$, most of the $m=1$ configurations
correspond to the fixed point $\phi _{i}(t)=2\pi/5$ ($i=1,...,5$),
whereas $m=0$ can emerge from five different fixed points on
$S_{1}$, namely, those of the type $\phi _{i}(t)=2\pi $ with $\phi
_{j}(t)=0$ for all $\,j\neq i$ ($i=1,...,5$). Likewise, $m=-1$ is
dominated by two sets of fixed points on $S_{1}$: five corresponding
to one link having undergone a $4\pi $ total slip, and ten
corresponding to two different links each having undergone a $2\pi $
slip. Note that, even for $m=1$ on $S_{1}$, there are many other
configurations different from the dominant ones mentioned above e.g.
$\phi _{i}=2\pi /5+2\pi,\,\phi _{j}=2\pi /5-2\pi,$ and $\,\phi
_{k}=2\pi /5$ for $k \neq i,j\,(i,j=1,...,5)$. However, in the limit
of large $\gamma$, those configurations involving many different,
mutually cancelling phase slips should have negligible probability.

To calculate the area of the basin of attraction for $m=1$, we
define a set of five orthonormal vectors $\hat{x}_i$ such that four
of them lie on $S_1$ and the fifth vector is perpendicular to $S_1$.
We define our origin on $S_1$ by shifting that of $S_0$ along
$\hat{x}_5$ by an amount $\varphi_1=2\pi/5$. The five vectors are
then given by: $\hat{x}_1=(1/\sqrt{ 2})(1,-1,0,0,0), \,
\hat{x}_2=(1/\sqrt{2})(0,0,1,-1,0),\, \hat{x}_3=(1/\sqrt{
20})(1,1,1,1,-4), \,\hat{x}_4=(1/2)(1, 1,-1,-1,0),\,
\hat{x}_5=(1/\sqrt{5} )(1,1,1,1,1)$.

%$\hat{x}_1=(1/\sqrt{2})\times(1\,\,\,\,\,\, -\hspace{-0.05 cm}1\,\,\,\,\,\, 0\,\,\,\,\,\, 0\,\,\,\,\,\,
%0), \, \hat{x}_2=(1/\sqrt{2})\times(0\,\,\,\,\,\, 0\,\,\,\,\,\, 1\,\,\,\,\,\,
%-\hspace{-0.05 cm}1\,\,\,\,\,\, 0),\, \hat{x}_3=(1/\sqrt{20})\times(1\,\,\,\,\,\, 1\,\,\,\,\,\, 1\,\,\,\,\,\, 1\,\,\,\,\,\, -\hspace{-0.05 cm}4),
%\\ \hat{x}_4=(1/2)\times(1\,\,\,\,\,\, 1\,\,\,\,\,\, -\hspace{-0.05 cm}1\,\,\,\,\,\, -\hspace{-0.05 cm}1\,\,\,\,\,\, 0),\, \hat{x}_5=(1/\sqrt{5})\times(1\,\,\,\,\,\,
%1\,\,\,\,\,\, 1\,\,\,\,\,\, 1\,\,\,\,\,\, 1)$.

To obtain the basin boundaries on the four-dimensional hypersurface
we next write the four independent $\phi _{i}$'s in terms of the
in-plane basis vectors $\hat{x}_{i}$'s ($i=1,...,4$) and transform
to spherical coordinates
$(r,\theta _{1},\theta _{2},\theta _{3})$. %we obtain:
%$\phi_1=rf_1(\theta_1,\theta_2,\theta_3),\phi_2=rf_2(\theta_1,\theta_2,\theta_3),\phi_3=rf_3(\theta_1,\theta_2,\theta_3),\phi_4=rf_4(\theta_1,\theta_2,\theta_3)$,
%where
%\begin{widetext}
%\begin{eqnarray}
%f_1(\theta_1,\theta_2,\theta_3)&=&\frac{\cos{\theta_1}\sin{\theta_2}\sin{\theta_3}}{\sqrt{2}}-\frac{\sqrt{5}}{6}\cos{\theta_2}\sin{\theta_3}+\frac{\cos{\theta_3}}{2}
%\nonumber \\
%f_2(\theta_1,\theta_2,\theta_3)&=&-\frac{\cos{\theta_1}\sin{\theta_2}\sin{\theta_3}}{\sqrt{2}}-\frac{\sqrt{5}}{6}\cos{\theta_2}\sin{\theta_3}+
%\frac{\cos{\theta_3}}{2}
%\nonumber \\
%f_3(\theta_1,\theta_2,\theta_3)&=&\frac{\sin{\theta_1}\sin{\theta_2}\sin{\theta_3}}{\sqrt{2}}-\frac{\sqrt{5}}{6}\cos{\theta_2}\sin{\theta_3}-
%\frac{\cos{\theta_3}}{2}
%\nonumber \\
%f_4(\theta_1,\theta_2,\theta_3)&=&-\frac{\sin{\theta_1}\sin{\theta_2}\sin{\theta_3}}{\sqrt{2}}-\frac{\sqrt{5}}{6}\cos{\theta_2}\sin{\theta_3}-
%\frac{\cos{\theta_3}}{2}
%\nonumber
%\end{eqnarray}
%\end{widetext}
Now, the potential energy is given by
$\mathcal{E}=-E_{J}\sum_{i}\cos {\phi _{i}}$ and the condition
$\partial \mathcal{E}/\partial r=0$ defines the boundary of the
basin of attraction. Shifting the origin back to $S_{0}$, the basin
boundary for $m=1$ on $S_{1}$ is then given by:
\begin{eqnarray}
f_{1}\sin {(rf_{1}+\varphi _{1})} &+&f_{2}\sin {(rf_{2}+\varphi
_{1})}  \nonumber
\\
+f_{3}\sin {(rf_{3}+\varphi _{1})} &+&f_{4}\sin {(rf_{4}+\varphi
_{1})}=0, \label{basin}
\end{eqnarray}
where the various $f_{k}=f_{k}(\theta _{1},\theta _{2},\theta _{3})$
%$f_2=f_2(\theta_1,\theta_2,\theta_3)$, $f_3=f_3(\theta_1,\theta_2,\theta_3)$ and $f_4=f_4(\theta_1,\theta_2,\theta_3)$
are obtained from a coordinate transformation. The probability
$P(1|1)$ to end up with $m=1$ having started from any point on
$S_{1}$ is given by the ratio $A_{1}/B_{1}$, where $A_{1}$ is the
area enclosed by the curve (\ref{basin}) on $S_{1}$ and $B_{1}$ is
the total area on $S_{1}$ subject to the initial constraints $\phi
_{i}(0)\in (-\pi ,\pi ]$. Using Monte Carlo, we obtain
$P(1|1)=0.03$. Similarly we also calculate $P(0|1)$ and $P(0|0)$ by
Monte Carlo, both yielding 0.94. Using this second result, the
symmetry between $m=1$ and $m=-1$, and the fact that
$P(1|0)+P(0|0)+P(-1|0)=1$, we can also obtain $P(1|0)=P(-1|0)=0.03$.
By contrast, the initial distributions are $Q(0)=0.6$ and
$Q(1)=Q(-1)=0.2$. Hence in the limit of large $\gamma $,
$P(1)/Q(1)=0.15$, which indicates a shrinking of the initial
distribution in favor of final zero winding number. Full scale
simulations based on Eqs. (\ref{phi})-(\ref{n}) yields for the same
ratio $0.14$. An exact agreement would require consideration of
infinitely many phase-slip histories.

In the passing, we would like to note that the above analysis holds
true strictly in the Josephson regime. Experiments with fully
merging independent BECs \cite{anderson} or the scenario of
quasi-condensates in BEC formation as envisaged by Zurek
\cite{zurek}, always go through an intermediate Josephson regime
when adjacent condensates start to overlap. However, a complete
study of the dynamics there would require going beyond the two-mode
Josephson Hamiltonian (\ref{hamiltonian}) for each junction. This is
clearly reflected in the outcome of experiments by Scherer {\it et
al.} \cite{anderson} where three independent BECs have been merged
to form stable vortices in the final BEC.

We thank A.~J.~Leggett and S.~Rajaram for a valuable discussions. P.G. wishes to thank Universidad Complutense de Madrid for its hospitality. This work has been
supported by NSF through Grant No. NSF-DMR-03-50842, by M.E.C. (Spain), Grant No. FIS2004-05120, and by the Ram\'{o}n Areces Foundation.

\end{document}